\documentclass[3p]{elsarticle}
\newcommand{\be}{\begin{equation}}
\newcommand{\ee}{\end{equation}}
\newcommand{\lb}[1]{\label{#1}}
\newcommand{\sty}{\scriptstyle}

\usepackage{txfonts}
\usepackage{epsfig}
\usepackage[obeyspaces]{url}
\usepackage{graphicx}
\begin{document}

\title{The Gompertz-Pareto Income Distribution}

\author[eq]{F.\ Chami Figueira}
\ead{fernanda@chami.com.br}

\author[ibge]{N.J.\ Moura Jr.}
\ead{newtonjunior@ibge.gov.br}

\author[if]{M.B.\ Ribeiro\corref{cor1}}
\ead{mbr@if.ufrj.br}

\cortext[cor1]{Corresponding author}
\address[eq]{Chemical School, Federal University of Rio de Janeiro-UFRJ,
             Rio de Janeiro, Brazil}
\address[ibge]{Brazilian Institute for Geography and Statistics-IBGE,
Rio de Janeiro,
Brazil}
\address[if]{Physics Institute, Federal University of Rio de Janeiro-UFRJ,
CxP 68532, Rio de Janeiro, RJ 21941-972, Brazil}

\begin{abstract}
This work analyzes the Gompertz-Pareto distribution (GPD) of personal
income, formed by the combination of the Gompertz curve, representing
the overwhelming majority of the economically less favorable part of
the population of a country, and the Pareto power law, which describes
its tiny richest part. Equations for the Lorenz curve, Gini coefficient
and the percentage share of the Gompertzian part relative to the total
income are all written in this distribution. We show that only three
parameters, determined by linear data fitting, are required for its
complete characterization. Consistency checks are carried out using
income data of Brazil from 1981 to 2007 and they lead to the conclusion
that the GPD is consistent and provides a coherent and simple
analytical tool to describe personal income distribution data.
\end{abstract}

\maketitle
\section{Introduction}\lb{intro}

The mathematical characterization of income distribution is an old
problem in economics. Vilfredo Pareto \cite{pareto} was the first
economist to discuss it in quantitative terms and it bears his name
the law he found empirically in which the tail of the cumulative
income distribution, formed by the richest part of the population of
a country, follows a power law pattern. Since then, the \textit{Pareto
power law} for income distribution has been verified to hold universally,
for various countries and epochs \cite{k80}. Despite the empirical
success of this law, the characterization of the lower income region,
representing the overwhelming majority of the population in any country,
remained an open problem. Various functions with an increasing number
of parameters were proposed by economists to represent the lower part,
or the whole, of the income distribution \cite{gklo06}. However, no
consensus emerged on what would be the most suitable way of
representing the whole income distribution of countries.

In the middle 1990s physicists became interested in problems which
until then were considered the exclusive realm of economists.
Econophysicists approached these problems in a data driven mode
\cite{dfss05,s09,s10,s10p}, that is, with none, or little,
consideration to the typical neoclassical economics mind-frame in
which axiomatic, some would say ideological \cite{f04,s04},
considerations take precedence over real data \cite{s10,b08}.
Ignoring this empirically flawed mindset
\cite{k01,k03,k04,kirman,mh04,o97,o09,c10}, efforts have been
made by econophysicists, helped later by a few non-representative
economists, to carefully study real data of economic nature. This
gave a new impetus to the income distribution problem due to an
emerging body of fresh results, as well as hints from statistical
physics on how it could be dynamically modeled \cite{b09}. 

Dr\u{a}gulescu and Yakovenko \cite{dy01,s05,yr09} advanced an
exponential type distribution of individual income similar to the
Boltzmann-Gibbs distribution of energy in statistical physics.
Chatterjee et al.\ \cite{ccm04} discussed an ideal gas model of
a closed economic system where total money and agents number are
fixed. Clementi et al.\ \cite{cgk07,cmgk08,cgk09} proposed the
k-generalized distribution as a descriptive model for the size
distribution of income, based on considerations of statistical
physics. Willis and Mimkes \cite{wm04} used log-normal and
Boltzmann distributions to argue in favor of a separate treatment
of waged and unwaged income. Moura Jr.\ and Ribeiro \cite{nm09}
showed that the Gompertz curve combined with the Pareto power law
provide a good descriptive model for the whole income distribution
and where the exponential appears as an approximation for the middle
portion of the individual income data. In this model the Gompertz
curve represents the overwhelming majority of the economically less
favorable part of the population, whereas the Pareto law describes
the richest part.

Regarding the related phenomenon of wealth distribution, related
because income and wealth are not the same quantity and, therefore,
should not be confused (see \cite{nm09} and \S \ref{br} below),
Solomon \cite{ss98} argued that a power-law wealth distribution
implies in Levy-flights returns, whereas Bouchaud and M\'{e}zard
\cite{bm00} reached a Pareto power-law wealth distribution in a
model containing exchange between individuals and random
speculative trading. Solomon and Richmond \cite{sr01} used a
generalized Lotka-Volterra model to show that the wealth
distribution among individual investors fulfills a power law,
Repetowicz et al.\ \cite{rhr05} studied a model of interacting
agents that allows agents to both save and exchange wealth, Coelho
et al.\ \cite{crbh08} revealed the existence of two distinct power
law regimes in wealth distribution, one for the super-rich and
another with smaller Pareto exponents for the top earners in income
data sets, and Scafetta et al.\ \cite{spw04} used a two-part function
stochastic model to discuss trade and investment dynamics of a
society stratified in two distinct classes (more on this in
\S\ref{br} below). Further references on income and wealth
distribution can be found in Yakovenko and Rosser \cite{yr09}, as
well as in \cite{nm09} and \cite{by10}.

The aim of this paper is to discuss further the model advanced by
Moura Jr.\ and Ribeiro \cite{nm09}. We show here that this combined
model, named as \textit{Gompertz-Pareto distribution} (GPD), provides
a simple way of modeling income distribution since it is formed by
simple functions and is fully characterized by three positive
parameters which can be determined by linear data fitting. We
discuss simple consistency tests in order to ascertain whether
or not the results produced by the model can recover basic features
of the original distribution, namely the Lorenz curves, the Gini
coefficients and the percentage share of the Gompertzian population
relative to the total income of the country. We conclude that the
GPD is consistent and provides a coherent and conveniently very
simple way of modeling income data. 

The GPD is a power-law tailed distribution and, as such, it is
likely to have a larger set of applications than just income
distribution. This is so because a very wide range of observed
phenomena in physical, biological and social sciences are known
to be described by power-law tailed distributions. For instance,
in physical sciences this is the case of galaxy distribution
\cite{ry98,gsp05}, relativistic cosmology
\cite{rib92,rib93,rib94,amr01,rib05,airs07} and turbulence
\cite{b96}. In human activities these distributions are found
in citation of scientific papers \cite{red98}, intensity of
wars \cite{rt98} and their military and civilian casualties
\cite{ari07,ari07b},
population of cities \cite{nr06} and stock prices \cite{aix03}. In
biological sciences, power-law tailed distributions were found in
botany \cite{n94}, genomics \cite{no08} and branching networks of
biological systems \cite{wb04}. Refs.\ \cite{n05} and \cite{k09}
provide several other examples of physical, biological and social
systems exhibiting power-law tailed distributions. The Gompertz
curve is known to be a good descriptor of population dynamics,
mortality rate and growth processes in biology \cite[see] [and
references therein]{nm09}. Therefore, a system whose distribution
is characterized by the combination of the Gompertz curve and a
power-law tail suggests that growth may possibly be one of the
main dynamical components of its underlying complex system dynamics.

The plan of the paper is as follows. In \S \ref{basic} we review
the basic equations for modeling income distribution data. Section
\ref{gpdis} presents the equations for the GPD of individual income
and extends the model to describe the most basic descriptive tools
used to measure income inequality, namely the Lorenz curve and the
Gini coefficient. We also discuss how the GPD has an exponential
type behavior in its middle part. Section \ref{br} applies the model
to the income data of Brazil from 1981 to 2007 and also presents
new results not available in \cite{nm09}. Consistency checks are
provided by re-obtaining the Lorenz curves, Gini coefficients and
the percentage share of the Gompertzian part of the distribution.
These are derived from the parameters of the model and compared
with the original, not model based, equivalent results. It is shown
that the results coming from the GPD are self-consisted. Section
\ref{con} ends the paper with the conclusions.

\section{Basic Equations}\lb{basic}

This section reviews very briefly the most essential quantities and
functions necessary for the analytical description of the individual
income distribution. We followed the comprehensive treatment provided
by Ref.\ \cite{k80}, although a slightly different notation and
normalization was adopted to match similar choices made in
Ref.\ \cite{nm09}.

Let us define $\mathcal{F}(x)$ to be the {\it cumulative income
distribution} giving the probability that an individual receives an
income less than or equal to $x$. Then the \textit{complementary
cumulative income distribution} $F(x)$ gives the probability that
an individual receives an income equal to or greater than $x$. It
follows from these definitions that $\mathcal{F}(x)$ and $F(x)$ are
related as follows, 
\be \mathcal{F}(x)+F(x)=100,
    \lb{ff}
    \ee
where the maximum probability was taken as 100\%. Here $x$ is a
normalized income, obtained by dividing the nominal, or real,
income values by some suitable nominal income average \cite{nm09}.
If both functions $\mathcal{F}(x)$ and $F(x)$ are continuous and
have continuous derivatives for all values of $x$, we have that,
\be d\mathcal{F}(x)/dx = f(x), \; \; \; \; dF(x)/dx=-f(x),
    \lb{c}
\ee
and
\be \int_0^\infty f(x)\:dx=100, \lb{norm1}
\ee
where $f(x)$ is the {\it probability density function of
individual income}, defined such that $f(x)\,dx$ is the
fraction of individuals with income between $x$ and $x+dx$.
The expressions above bring about the following results,
\be \mathcal{F}(x) - \mathcal{F}(0) = \int_0^x f(w) \: dw,
    \lb{3}
\ee
\be F(x) - F(\infty) = \int_x^\infty f(w) \: dw.
    \lb{4}
\ee
The boundary conditions below approximately apply to our problem,
\be \left\{ \begin{array}{lclcl}
    \mathcal{F}(0) & = & {F}(\infty) & \cong & 0, \\
    \mathcal{F}(\infty) & = & {F}(0) & \cong & 100.
    \end{array}
    \right.
   \lb{condi}
\ee
Clearly both $\mathcal{F}(x)$ and $F(x)$ vary from 0 to 100. It is
simple to see that these conditions, together with the definitions
(\ref{c}), lead the normalization (\ref{norm1}) to be written as
follows,
\be \int_0^{100} d \mathcal{F}=-\int_{100}^0 dF =
    \int_0^\infty f(x)\:dx=100.
    \lb{norm2}
\ee

The average income of the whole population may be written as,
\be
  \langle x \rangle =\frac{\displaystyle \int_0^\infty x \: f(x) \:
  dx}{\displaystyle \int_0^\infty f(x)\:dx}
  =\frac{1}{100}\int_0^\infty x \: f(x) \: dx,
  \lb{avg1}
\ee
whereas the first-moment distribution function $\mathcal{F}_1(x)$
is given by,
\be \mathcal{F}_1(x)=100 \: \frac{\displaystyle \int_0^x w
    \: f(w) \: dw} {\displaystyle \int_0^\infty w \: f(w) \: dw}
    =\frac{1}{\langle x \rangle} \int_0^x w \: f(w) \: dw.
    \lb{f1}
\ee
Thus, $\mathcal{F}_1(x)$ varies from 0 to 100 as well.

One of the most common tools to discuss income inequality is the
\textit{Lorenz curve}, comprising of a 2-dimensional curve whose
x-axis is the
proportion of individuals having an income less than or equal to
$x$, whereas the y-axis is the proportional share of total income
of individuals having income less than or equal to $x$. In other
words, the horizontal coordinate of the Lorenz curve represents the
fraction of population with income below $x$ and the vertical
coordinate gives the fraction of total income of the population
receiving income below $x$, as a fraction of the total income of
this population \cite[p.\ 30]{k80}.
Analytically, the cumulative income distribution $\mathcal{F}(x)$
given by equation (\ref{3}) and boundary condition (\ref{condi})
defines the \textit{x-axis of the Lorenz curve}, that is,
\be \mathcal{F}(x) = \int_0^x f(w) \: dw,
    \lb{a}
\ee
whereas the \textit{y-axis of the Lorenz curve} is defined by the
first-moment distribution function $\mathcal{F}_1(x)$ given by
equation (\ref{f1}). The curve is usually represented in a unit
square, but due to the normalization (\ref{norm1}) above, here
the square where the Lorenz curve is located has area equal to
$10^4$.

The Lorenz curve allows us to define another commonly used index
to measure the inequality of the income distribution, the \textit{
Gini coefficient}. This is constructed with the ratio of the area
between the egalitarian line, defined as the diagonal connecting
points (0,0) and (100,100), and the Lorenz curve to the area of
the triangle beneath the egalitarian line \cite[pp.\ 32, 71]{k80}.
The expression of this coefficient under the currently adopted
normalization may be written as,
\be Gini = 1- 2\times 10^{-4} \int_0^{100} \mathcal{F}_1 \:
         d \mathcal{F}
       = 1- 2\times 10^{-4} \int_0^\infty \mathcal{F}_1(x)
         \: f(x) \: dx.
    \lb{gini}
\ee

\section{The Gompertz-Pareto Income Distribution}\lb{gpdis}

It was advanced in Ref.\ \cite{nm09} that the complementary
cumulative income distribution is well described by two components.
The first, representing the overwhelming majority of the population
($\sim$ 99\%), is given by a \textit{Gompertz curve}, whereas the
second, representing the richest tiny minority ($\sim$ 1\%), is
described by the \textit{Pareto power law}. Then, the complementary
cumulative distribution yields, 
\be F(x)= \left\{ \begin{array}{lcllc}
          G(x) & = & e^{ \displaystyle e^{(A-Bx)} },
	  & \; \; ( \: 0 \le x < x_{\sty t}), 
	  & \; \mbox{(Gompertz)} \\
	  P(x) & = & \beta \; x^{\displaystyle - \alpha},
	  & \; \; (x_{\sty t} \le x \le \infty),
	  & \; \mbox{(Pareto)} \\
	  \end{array}
	  \right.
          \lb{distro}
\ee
and the cumulative income distribution may be written as
below,
\be \mathcal{F}(x)= \left\{
    \begin{array}{ll}
      \mathcal{G}(x)=100-e^{ \displaystyle e^{(A-Bx)} },
      & \; \; ( \: 0 \le x < x_{\sty t}), \\
      \mathcal{P}(x)=100-\beta \; x^{\displaystyle - \alpha},
      & \; \; (x_{\sty t} \le x \le \infty). \\
    \end{array}
    \right.
    \lb{disto1}
\ee
Here $x_{\sty t}$ is the income value threshold of the Pareto
region. It follows from these equations that the probability density
income distributions of both components may be written
according to the expressions below,
\be f(x)= \left\{ \begin{array}{lclc}
          g(x) & = & B \; e^{(A-Bx)} \; 
          e^{ \displaystyle e^{(A-Bx)} },
	  & \; \; ( \: 0 \le x < x_{\sty t}),  \\
          p(x) & = & \alpha \; \beta \; x^{^{\scriptstyle
	  -(1+\alpha )}}, & \; \; (x_{\sty t} \le x \le
	  \infty).  \\
	  \end{array}
	  \right.
          \lb{distro2}
\ee

This distribution is seemingly characterized by five parameters: $A$, $B$,
$x_{\sty t}$, $\alpha$, $\beta$. There are, however, two additional
constraints and one restriction which reduce the parametric freedom
of the distribution. Firstly, the boundary conditions (\ref{condi})
determine the value of $A$. Indeed, we have that,
\be F(0)=100 
    \; \; \Longrightarrow \; \; A=\ln \left( \ln 100 \right).
    \lb{ic}
\ee
Secondly, the normalization (\ref{norm1}) of the probability density,
written as,
\be
 \int_0^{x_{\sty t}} B \; e^{(A-Bx)} \; e^{ \displaystyle e^{(A-Bx)} 
  } dx + \int_{x_{\sty t}}^\infty \alpha \; \beta \;
 x^{^{\scriptstyle  -(1+\alpha )}} dx=100,
 \lb{norm}
\ee
and the continuity of the functions (\ref{distro}) across the frontier
between the Gompertz and Pareto regions, defined as $x=x_t$, both
lead to the determination of $\beta$ by means of the following constraint
equation,
\be
   \beta= {(x_{\sty t})}^{\displaystyle \alpha} \;
          e^{ \displaystyle e^{(A-Bx_{\sty t})} }.
   \lb{vinc}
\ee
In addition, considering eqs.\ (\ref{avg1}) and (\ref{distro2}),
it is simple to show that the average income of the whole population
in the GPD may be written as follows,
\be
  \langle x \rangle =
  \frac{1}{100} \left[ \mathcal{I}(x_t) + \frac{\alpha \:
  \beta}{(\alpha -1)} {x_{\sty t}}^{-(\alpha-1)} \right],
      \lb{avg}
\ee
where $\mathcal{I}(x)$ is given by the following, numerically
solvable, integral,
\be \mathcal{I}(x)
    \equiv \int_0^x w \, g(w) \, dw=
    \int_0^x w \: B \: e^{(A-Bw)} \;
    e^{ \displaystyle e^{(A-Bw)}} dw. 
    \lb{I}
\ee
Clearly the average in eq.\ (\ref{avg}) will only converge if
\be \alpha > 1. \lb{am1} \ee
As discussed in Ref.\ \cite{nm09}, although extremely rich
individuals do exist, there are limits to their wealth and, hence,
the average income cannot increase without bound. 

Summarizing, \textit{the Gompertz-Pareto distribution 
is fully characterized by three parameters under the following
restrictions},
\be
  \left\{ \begin{array}{rl}
    \alpha  > 1, \\
    x_{\sty t}  > 0, \\
    B  > 0.
    \end{array}
    \right.
    \lb{rts}
\ee
These parameters can be determined directly from observed data, that
is, from a sample of $n$ observed income values $x_j$, such that,
\be
  \{x_j\}: (j=1,\ldots,n), \, (x_1=x_{\mathrm{min}}).
  \lb{obs}
\ee
Inasmuch as both equations (\ref{distro}) can be linearized, we can
determine the unknown parameters by linear data fitting. It should be
noted, however, that minimal 3-parameters fits also appear in other
models of income and wealth distribution, like in Scafetta et al.\
\cite{spw04} and Banerjee and Yakovenko \cite{by10}.

\subsection{Exponential Approximation}\lb{exp}

It is known that the middle section of the income distribution
data from various countries can be modeled by exponential-type
functions \cite{s05,yr09,ccm04,cgk07, cgk09,spw04}. Under
suitable approximation the GPD does allow for
this empirical feature to hold \cite{nm09}. For large values of $x$
the term $Bx$ dominates over the parameter $A$ in the first
equation (\ref{distro}), allowing us to write that $G(x)
\approx e^{\displaystyle e^{-Bx}}$. In addition, when $e^{-Bx}
< 1$, the Taylor expansion below holds,
\be e^{\displaystyle e^{-Bx}}=1+e^{-Bx} + \frac{1}{2}
    {\left(e^{-Bx}\right)}^2+\frac{1}{6} {\left(e^{-Bx}\right)}^3
    + \ldots
    \lb{apx}
\ee
The density $g(x)$ in eq.\ (\ref{distro2}) can also be similarly
approximated and, therefore, we can write the following exponential
approximations for the middle and upper sections of the GPD, 
\be \left\{ \begin{array}{ll}
    G(x) & \approx 1+  e^{-Bx}, \\ 
    g(x) & \approx B \; e^{-Bx}.
    \end{array}
    \right.
   \lb{gex}
\ee
These approximations hold only in the Gompertzian part of the
distribution, i.e., for $x<x_t$.

\subsection{The Lorenz Curve}\lb{lrz} 

As discussed above, the first-moment distribution function
$\mathcal{F}_1(x)$ given by equation (\ref{f1}) defines the
y-axis of the Lorenz curve, whereas the cumulative income
distribution function $\mathcal{F}(x)$ given by eq.\ (\ref{a})
defines the x-axis. Applying equations (\ref{distro2}) to these
definitions and considering eqs.\ (\ref{ic}), (\ref{avg}) and (\ref{I}),
the axes of the Lorenz curve for the GPD yield,
\be \mathcal{F}(x)= \left\{ \begin{array}{lr}
    100-e^{\displaystyle e^{(A-Bx)}},
    & (0 \le x<x_t), \\
    100-e^{\displaystyle e^{(A-Bx_t)}}
    - \beta \left( x^{\, -\alpha} - {x_t}^{-\alpha} \right),
    & \; \; \; (x_t\le x<\infty),
    \end{array}
    \right.
    \lb{lorenz-totalX}
\ee
and
\be \mathcal{F}_1(x)= \left\{ \begin{array}{lr}
    \displaystyle \frac{\mathcal{I}(x)}{ \langle x \rangle}
    \,, & (0<x<x_t), \vspace{4mm} \\
    100+\displaystyle 
    \frac{\displaystyle \alpha \, \beta}{\displaystyle
    (1-\alpha)} \; \frac{x^{(1-\alpha)}}{\langle
    x \rangle}, & \; \; \; (x_t\le x<\infty;).
    \end{array}
    \right.
    \lb{lorenz-totalY}
\ee

\subsection{Gini Coefficient}\lb{gn}

The Gini coefficient as defined by equation (\ref{gini}) must
now take into consideration the results appearing in equations
(\ref{distro2}) and (\ref{lorenz-totalY}). Thus, in the GPD,
equation (\ref{gini}) becomes,
\be Gini=1-2\times 10^{-4} \left\{ \frac{B}{\langle x \rangle}
         \int\limits_0^{x_t} \mathcal{I}(x)\: e^{(A-Bx)}
	 e^{ \displaystyle e^{(A-Bx)} } dx + 100\:\beta\:{x_t}^{-\alpha}
	 +\frac{\alpha^2 \beta^2 \: {x_t}^{(1-2\alpha)}}{\langle x
	 \rangle (\alpha-1)(1-2\alpha)} \right\}.
	 \lb{gini2}
\ee

\section{Application to the Brazilian Data: 1981 - 2007}\lb{br}

The income distribution of Brazil from 1978 to 2005 was
detailed studied by Moura Jr.\ and Ribeiro \cite{nm09}, where
it was shown that the GPD provides a good representation for the
Brazilian income data. All parameters of this distribution were
fitted to this time span, although it became clear that the results
for 1978 and 1979 were prone to large errors resulting from probable
inconsistencies in the original sample. Due to this, here we shall
disregard the data for these two years, but include
previously unpublished results for 2006 and 2007. Table~\ref{tab1}
presents the parameters of the Gompertz-Pareto income distribution
for Brazil from 1981 to 2007, as well as values for $u$, the
\textit{percentage share of the Gompertzian part of the income
distribution}, and the Gini coefficient.
\begingroup
\begin{table*}[!htbp]
\caption{Parameters of the GPD for the income data of Brazil. The
         results from 1981 to 2005 were first shown in Ref.\ \cite{nm09},
         whereas for 2006 and 2007 are new. The theoretically predicted
         value $A=1.52718$ in eq.\ (\ref{ic}) was found with
         a maximum discrepancy of 2.15\%. $B$, $x_t$ and $\alpha$ were
         obtained by linear data fitting and $\beta$ was found by means
         of the constraint equation (\ref{vinc}), where the theoretical
         result for $A$ was used. Lorenz curves were generated from the
         data for each year, allowing the calculation of the Gini
         coefficients. Once $x_t$ was found, it became possible to
         determine $u$, the percentage share of the Gompertz part of the
         income distribution, directly from the data. See \cite{nm09} for
         details on these calculations. The last two columns on the right
         show the results for the Gini coefficient and the percentage
         share of the Gompertzian segment calculated by using $\alpha$,
         $x_t$ and $B$ in equations (\ref{gini2}) and (\ref{u2}). These
         calculated values are denoted as $Gini^*$ and $u^*$ in
         order to differentiate them from the original values $Gini$ and
         $u$ obtained without assuming the GPD. Errors for $Gini^*$ and
         $u^*$ were estimated by quadratic propagation and are provided
         here just as a general indication of uncertainties since we are
         not dealing with a tightly controlled experimental environment.
         Comparison of both Gini coefficient values show that the
         original $Gini$ results fall under the calculated errors of
         $Gini^*$. If we dismiss these uncertainties, we note that the
         values of $Gini^*$ have a maximum discrepancy of 7\% to the
         original $Gini$ ones. Similarly, $u^*$ was calculated by means of
         equation (\ref{u2}) and uncertainties were obtained by quadratic
         propagation where we allowed for a 2.15\% uncertainty in $A$
         (see above). If one dismisses the uncertainties in $u^*$, one
         can verify that the discrepancies between $u$ and $u^*$ are not
         higher than 6\%, a result which indicates a good consistency
         between the GPD and Brazil's income data.\label{tab1}}
\begin{center}
\begin{tabular}{ccccccccc}
\hline\noalign{\smallskip}
year & $B$ & $x_{\sty t}$ & $\alpha$ & $\beta$ & $Gini$ & $u$ & $Gini^*$ & $u^*$ \\ 
\noalign{\smallskip}\hline\noalign{\smallskip}
1981 &$0.342\pm0.016$&7.533&$2.839\pm0.109$ & $438\pm98$ & $0.574$ & $87.7$ & $0.613\pm0.088$ & $82.5\pm5.1$ \\
1982 &$0.342\pm0.015$&7.473&$2.677\pm0.057$ & $312\pm38$ & $0.581$ & $87.1$ & $0.615\pm0.049$ & $82.0\pm3.2$ \\
1983 &$0.330\pm0.010$&6.910&$2.636\pm0.047$ & $261\pm25$ & $0.584$ & $85.5$ & $0.611\pm0.039$ & $81.6\pm2.8$ \\
1984 &$0.332\pm0.013$&7.388&$2.839\pm0.109$ & $434\pm96$ & $0.576$ & $87.2$ & $0.611\pm0.087$ & $82.4\pm5.1$ \\
1985 &$0.329\pm0.010$&7.490&$2.656\pm0.052$ & $311\pm34$ & $0.589$ & $85.8$ & $0.614\pm0.044$ & $81.9\pm3.0$ \\
1986 &$0.344\pm0.013$&7.112&$2.567\pm0.034$ & $229\pm17$ & $0.580$ & $85.2$ & $0.615\pm0.031$ & $81.6\pm2.5$ \\
1987 &$0.343\pm0.016$&7.626&$2.724\pm0.070$ & $354\pm52$ & $0.592$ & $85.9$ & $0.615\pm0.059$ & $82.2\pm3.7$ \\
1988 &$0.324\pm0.015$&8.140&$2.874\pm0.122$ & $576\pm149$& $0.609$ & $85.4$ & $0.614\pm0.102$ & $82.6\pm5.8$ \\
1989 &$0.317\pm0.010$&7.856&$2.777\pm0.086$ & $448\pm81$ & $0.628$ & $82.5$ & $0.612\pm0.072$ & $82.3\pm4.3$ \\
1990 &$0.335\pm0.016$&8.074&$2.636\pm0.047$ & $335\pm36$ & $0.605$ & $85.9$ & $0.618\pm0.044$ & $81.8\pm3.0$ \\
1992 &$0.364\pm0.019$&7.635&$2.636\pm0.047$ & $283\pm30$ & $0.578$ & $87.0$ & $0.619\pm0.044$ & $81.8\pm2.9$ \\
1993 &$0.330\pm0.008$&7.674&$2.567\pm0.034$ & $270\pm19$ & $0.599$ & $84.1$ & $0.616\pm0.030$ & $81.6\pm2.4$ \\
1995 &$0.333\pm0.012$&7.887&$2.777\pm0.086$ & $432\pm78$ & $0.596$ & $85.9$ & $0.615\pm0.072$ & $82.3\pm4.3$ \\
1996 &$0.347\pm0.020$&8.163&$2.749\pm0.077$ & $421\pm71$ & $0.598$ & $86.7$ & $0.619\pm0.068$ & $82.1\pm4.1$ \\
1997 &$0.338\pm0.016$&7.935&$2.617\pm0.043$ & $310\pm30$ & $0.598$ & $86.1$ & $0.618\pm0.040$ & $81.8\pm2.8$ \\
1998 &$0.326\pm0.009$&7.628&$2.677\pm0.057$ & $338\pm40$ & $0.597$ & $84.5$ & $0.614\pm0.048$ & $81.9\pm3.2$ \\
1999 &$0.331\pm0.013$&7.811&$2.777\pm0.086$ & $426\pm77$ & $0.590$ & $86.0$ & $0.614\pm0.072$ & $82.3\pm4.3$ \\
2001 &$0.335\pm0.011$&7.774&$2.724\pm0.070$ & $375\pm55$ & $0.592$ & $85.2$ & $0.615\pm0.059$ & $82.1\pm3.7$ \\
2002 &$0.339\pm0.015$&7.878&$2.777\pm0.086$ & $424\pm77$ & $0.586$ & $86.4$ & $0.615\pm0.073$ & $82.3\pm4.3$ \\
2003 &$0.333\pm0.009$&7.374&$2.777\pm0.086$ & $381\pm67$ & $0.579$ & $85.4$ & $0.612\pm0.070$ & $82.3\pm4.2$ \\
2004 &$0.342\pm0.015$&7.653&$3.104\pm0.226$ & $775\pm358$& $0.582$ & $87.2$ & $0.611\pm0.175$ & $83.1\pm9.7$ \\
2005 &$0.326\pm0.009$&7.403&$2.839\pm0.109$ & $444\pm97$ & $0.580$ & $86.2$ & $0.610\pm0.087$ & $82.4\pm5.0$ \\
2006 &$0.327\pm0.014$&7.910&$3.749\pm0.561$ &$3295\pm3824$&$0.581$ & $87.9$ & $0.605\pm0.408$ & $84.2\pm22.4$ \\
2007 &$0.334\pm0.009$&6.934&$2.839\pm0.109$ & $385\pm82$ & $0.572$ & $85.7$ & $0.608\pm0.084$ & $82.3\pm4.9$ \\
\noalign{\smallskip}\hline
\end{tabular}
\end{center}
\end{table*}
\endgroup

At this point it is important to note that the Gini coefficients
can be obtained without any assumption regarding the shape and
functional form of the income distribution, that is, they can be 
obtained independently of the GPD. Similarly, although $x_t$ is
used as a cut-off income value necessary to obtain $u$, its
evaluation does not require information about the shape and form
of the distribution and, hence, it is also model independent.
These \textit{original} values for $Gini$ and $u$ obtained
directly from the data, are shown unmarked in columns 6 and 7
from left to right in Table~\ref{tab1}. These remarks make it
possible to check the consistency of the Gompertz-Pareto
representation of the Brazilian income distribution by rebuilding
the Lorenz curves for each year, re-obtaining the Gini coefficients
by means of equation (\ref{gini2}) and comparing with the original
ones. 

Similar calculation is possible to do with $u$ once we note that,
by definition, we may write the following equation,
\be u=\mathcal{F}_1(x_t).
    \lb{u}
\ee
Considering eqs.\ (\ref{vinc}) and (\ref{lorenz-totalY}), we reach
an expression linking the percentage share of the lower income class
with the parameters of the GPD. It may be written as follows,
\be u  =100-\frac{\alpha}{(\alpha-1)}\frac{x_t}{\langle x \rangle}
       e^{ \displaystyle e^{(1.52718-Bx_t)}}.
       \lb{u2}
\ee

Figure \ref{fig1} shows the
Lorenz curves for Brazil obtained from the GPD using the values of
$\alpha$, $x_t$ and $B$ provided in Table~\ref{tab1} in equations
(\ref{lorenz-totalX}) and (\ref{lorenz-totalY}). Vertical and
horizontal error bars obtained by standard error propagation
techniques are provided as a general indication of uncertainties.
The plots show that the curves are consistent with the behavior one
would expect of the Lorenz curves and compare satisfactorily with
the original ones presented in Ref.\ \cite{nm09}.
\begin{figure}[htbp]
\begin{center}
\includegraphics[scale=1.66]{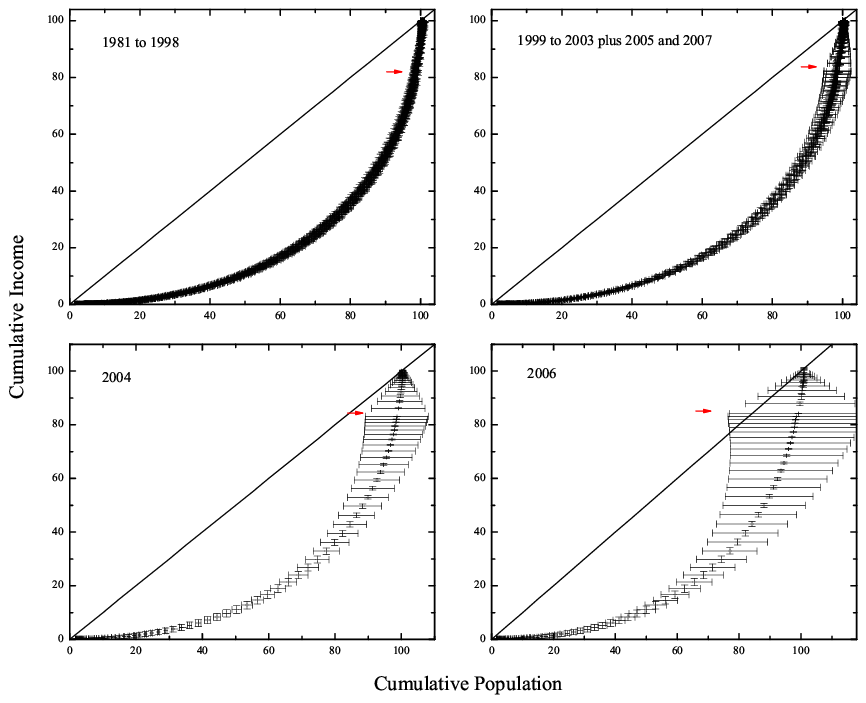}
\caption{Lorenz curves for Brazil from 1981 to 2007 obtained by
         using the GPD parameters of Table~\ref{tab1} in equations
	 (\ref{lorenz-totalX}) and (\ref{lorenz-totalY}). The small
         arrows indicate the approximate point of transition from the
         Gompertz region to the power-law regime. Vertical
	 and horizontal error bars represent uncertainties calculated
         by standard error propagation techniques. They are divided in
         two groups according to how well similar curves collapse to a
         single curve. The top left graph shows the curves from 1981
         to 1998, whereas the top right plot presents the Lorenz curves
         from 1999 to 2007, except 2004 and 2006 which are both shown
         separately at the bottom. The plots themselves show clearly
         that, excluding 2004 and 2006, all other curves fall in 
         two distinct groups, since the collapsed curves become very
         well defined. The Brazilian Lorenz curves present a remarkable
         stability in their respective time frames, even considering
         the hyperinflation period, which is included in the top left
         plot. The graphs for 2004 and 2006 are shown individually at
         the bottom because in these volatility is highest. This is a
         consequence of the fact that the Brazilian agency responsible
         for collecting income data carried out a much more restricted
         sampling in those years, resulting in much shorter Pareto tails
	 and, hence, higher fluctuations, as compared to the other
	 years. Since the Gompertz curve is a double exponential, larger
	 fluctuations are greatly amplified at the middle upper range of
	 the Gompertzian section of the distribution.}\lb{fig1}
\end{center}
\end{figure}

The results for the Gini coefficient and the percentage share of
the Gompertzian part obtained by using the parameters $\alpha$,
$x_t$ and $B$ of Table~\ref{tab1} in equations (\ref{gini2}) and
(\ref{u2}) are shown at the last two columns on the right in
Table~\ref{tab1}. These were \textit{calculated} by assuming the
GPD and are shown as $Gini^*$ and $u^*$. Uncertainties were also
calculated by standard error propagation techniques, but should
not be viewed at their face values, but just as general indications
of error margins since we are not dealing with experimental
errors stemming from experimental devices in carefully controlled
environments available in laboratories where measurement limitations
can be precisely determined. However, one can see by comparing $Gini$
with $Gini^*$ and $u$ with $u^*$ that in general the calculations
recover both quantities, indicating an overall consistency between the
GPD and the individual income data of Brazil from 1981 to 2007.

Figure \ref{fig5} shows a plot for both original $Gini$ and
calculated $Gini^*$ coefficients appearing in Table \ref{tab1}.
One can verify a general agreement between both results, indicating
a good consistency between the GPD and the Brazilian personal
income data in the studied time span. A better comparison is
shown in Fig.\ \ref{fig6} where the curves were zoomed in and error
bars removed for better clarity. It is clear from this plot that our
calculated $Gini^*$ values were systematically overestimated as
compared the original $Gini$. However, this difference is small,
having a maximum discrepancy of 7\%. That might be a result of a
possible statistical bias, probably present in the original
estimation of the GPD parameters. In any case, one can verify a
general agreement in the evolving tendency of the two curves. From
1983 to 1993 there are visible high fluctuations in the original
Gini coefficients, a period which is within the high inflationary
period Brazil went through by the end of the last century. In fact,
the peak of this period is 1989, when Brazil suffered from
hyperinflation reaching almost three digits per month. After 1994,
the year when inflation came to an abrupt end, the two lines tend
to follow each other with a systematic, but stable, difference.
\begin{figure}[htbp]
\begin{center}
\includegraphics[angle=-90,scale=0.46]{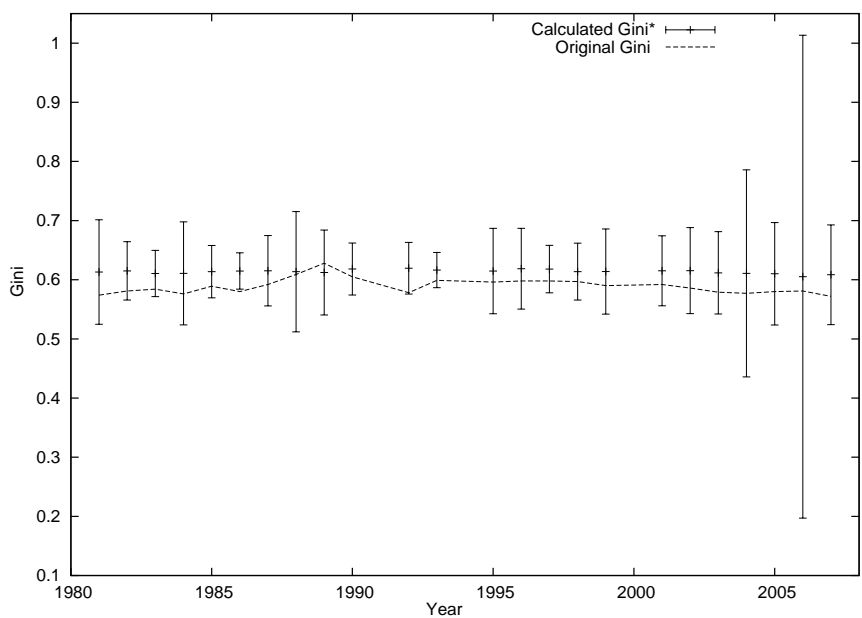}
\caption{Plots comparing the original Gini coefficients with the
        calculated ones in Table \ref{tab1}. One can see a general
        agreement between both sets of points, indicating consistency
        between the GPD and Brazil's personal income data in the
        period analyzed here.}\lb{fig5}
\end{center}
\end{figure}
\begin{figure}[htbp]
\begin{center}
\includegraphics[angle=-90,scale=0.46]{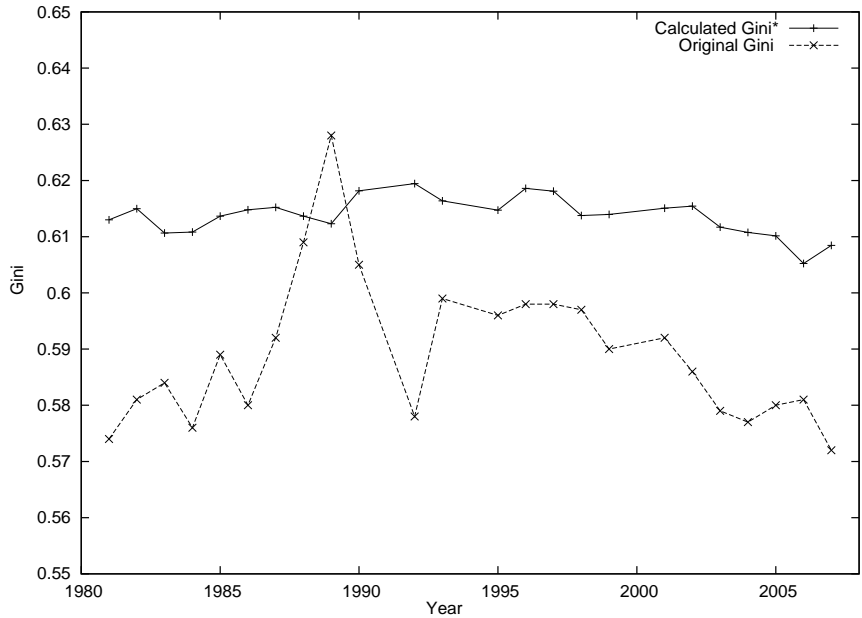}
\caption{This graph shows the same results of Fig.\ \ref{fig5},
         but zoomed in and without error bars for better clarity.
	 High fluctuations can be seen from 1983 to 1993 in the
	 original $Gini$ values, a period which coincides with
         very high inflation, peaking with hyperinflation in 1989,
         the highest peak of the lower curve. The plot also shows
         a systematic difference between both lines for most of
         the studied time span, varying mostly from 0.02 to 0.03
         and reaching its maximum of 0.041 in 1992 which is within
         the strong inflationary period Brazil experienced at that
         time. Despite this systematic difference, which might be
         a result of some statistical bias present in the original
         determination of the GPD parameters, one can observe a
         general consistency between both curves, especially if we
         bear in mind that this discrepancy does not go higher than
         a 7\%, a value which could possibly be taken as the upper
         limit of this possible bias.}\lb{fig6}
\end{center}
\end{figure}

The results for the percentage share of the Brazilian population
whose income is inside the Gompertzian part of the distribution are
shown in Fig.\ \ref{fig7}. There we can see again a general consistency
between the original $u$ values with the calculated $u^*$ of Table
\ref{tab1}. Figure \ref{fig8} shows the same results, but zoomed in and
without error bars. Similarly to the Gini coefficients, one can verify
a systematic difference between both lines, but now the calculated
values $u^*$ are underestimated as compared to the original ones. We can
again see high fluctuations in the original values from 1988 to 1994, a
period within the high inflationary era in Brazil. The deepest valley
occurs in 1989, the year of highest hyperinflation in Brazil.
Nevertheless, the two curves tend to evolve in a similar fashion, also
featuring an approximately stable discrepancy whose maximum is 6\%.
\begin{figure}[htbp]
\begin{center}
\includegraphics[angle=-90,scale=0.46]{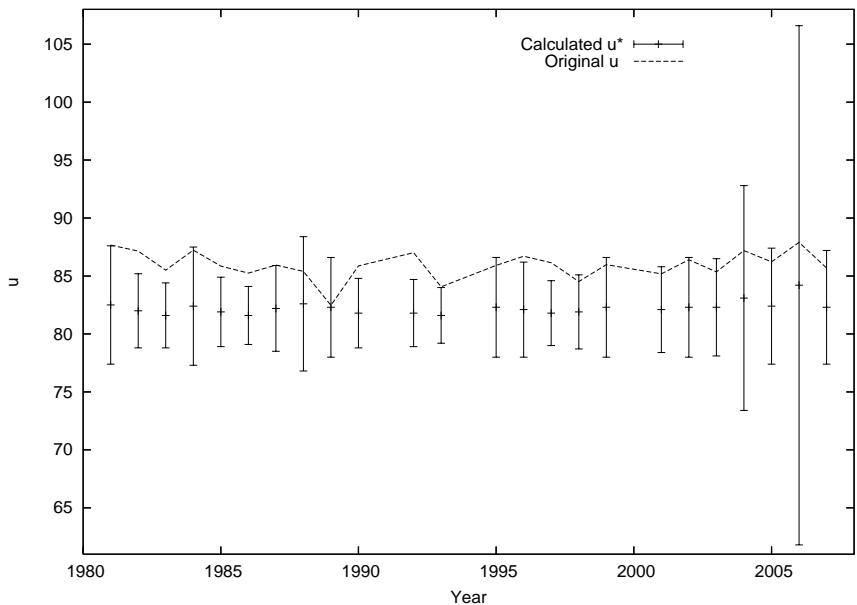}
\caption{This graph shows the evolution of $u$, the percentage share of
         the total income of the Gompertzian part of the distribution,
	 originally obtained without assuming the GPD, as compared with
         the calculated ones listed as $u^*$ in Table \ref{tab1} and
         obtained using the fitted GPD parameters. Similarly to the case
         of the Gini coefficients, one can see a general consistency
         between both results, although a systematic discrepancy is also
         present (see Fig.\ \ref{fig8}).}\lb{fig7}
\end{center}
\end{figure}
\begin{figure}[htbp]
\begin{center}
\includegraphics[angle=-90,scale=0.46]{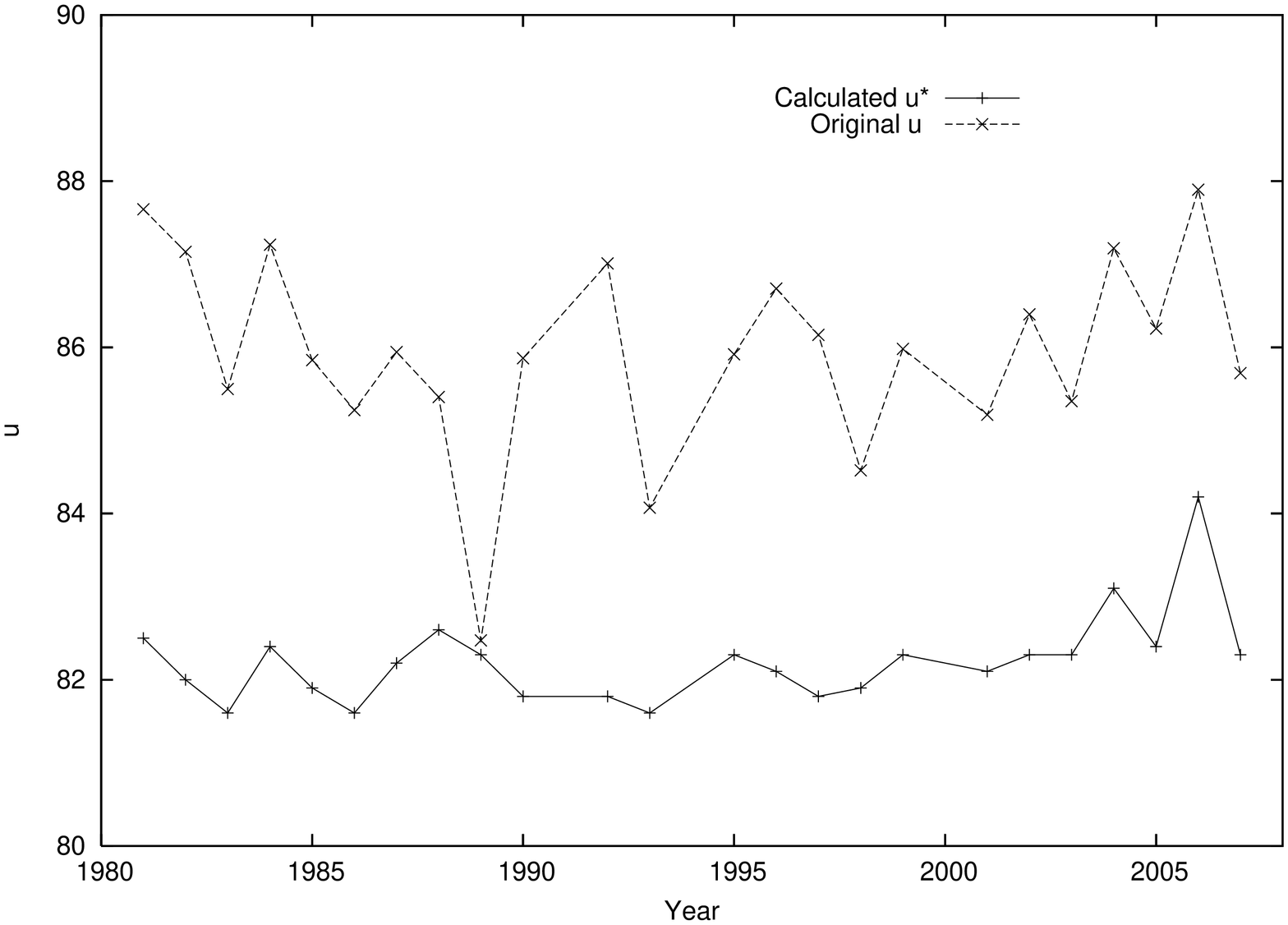}
\caption{This graph shows the same results of Fig.\ \ref{fig7}, but
         zoomed in and with error bars removed for better clarity.
	 We can verify that there is a systematic underestimation of
	 the calculated results for $u^*$ as compared to the original
         ones listed as $u$ in Table \ref{tab1}.
	 This discrepancy has, however, a maximum value of 6\%, being,
	 therefore, very close to the one found in the Gini coefficients
	 (see Fig.\ \ref{fig6}). One can also see high fluctuations
	 in the original results during the inflationary period of
	 Brazil. However, here these fluctuations seem to be
	 restricted to the somewhat shorter period lasting from 1988
	 to 1994. The deepest valley occurs in 1989, the year when
	 Brazil was hit by hyperinflation. Since this systematic
	 discrepancy is small and mostly stable, the results indicate
	 that overall the GDP provides a good and consistent way of
	 modeling income distribution data.}\lb{fig8}
\end{center}
\end{figure}

As final comments, one may ask if a combined two-part function is
more appropriate to describe income distribution rather than a
single function, no matter how complicated. It was argued in
Ref.\ \cite{nm09} that from an econophysical viewpoint the paramount
objective of an accurate empirical characterization of income
distribution is to reveal the underlying dynamics of this system
and its governing differential equations. On this point one should
mention the model advanced by Scafetta et al.\ \cite{spw04}
\cite[see also][]{spw02,spw04b} where the distribution of
\textit{wealth}, not \textit{income}, can be explained by a
two-part function, where the low to medium range is
fitted to the gamma function and the high wealth is fitted to the
Pareto power-law. If the less wealthy has in trade the origin of
their resources, with trade being statistically biased in favor of
the poor, and the rich obtain their resources from investment,
then the model reproduces the stratification of society into a
small upper class comprising about 1\% of the population and the
remaining 99\% forming a large middle class together with a poor
class. So, two functions mean two different, but inter-related,
dynamics: the gamma function would represent returns in trade and
the Pareto power-law returns in investment. So, the less wealthy
trade with an advantage their only low-return resource, their
own labor.\footnote{We are grateful to a referee for pointing
this out.}

To reach these conclusions, Ref.\ \cite{spw04} developed a stochastic
model built upon some economic concepts which may provide useful in
further studies of the dynamics of income distribution. Thus, wealth
should not be confused with income, since, although related, the
former comprises all assets and liabilities of a person reported
at a certain moment, e.g., at the person's death, whereas income is
the quantity of money, or its equivalent, a person receives in a
certain period of time in exchange for sale of goods or property,
services, labor or profit from financial investments. So, similarly
to \cite{nm09}, it seems reasonable to state that income is a flux
of money, or its equivalent, per time unit, whereas wealth could be
thought of as income less consumption integrated over a period of
time plus a constant representing assets obtained in a previous
time period. In addition, Scafetta et al.\ \cite{spw04} define investment
as ``any act that creates or destroys wealth'' and trade as ``any
type of economic transaction.'' Accordingly, in a trade transaction
the total wealth is conserved and the rich receive their returns
from investments as they own the means of large production. They
conclude by arguing that this trade bias in favor of the poor is
not only possible, but necessary so that society is stabilized in
order to avoid the catastrophic situation where the entire wealth
of the society becomes concentrated in the hands of very few
extremely wealthy people.

Therefore, a two-part function may provide important hints to the
underlying dynamics of income distribution, hints on the relationship
between the upper and lower sections of the distribution function
which would otherwise remain hidden if one were to use a single
distribution function. This seems specially true when one considers
that society is formed by economically distinct classes that may 
be better represented by distinct functions, which in turn possess
distinct, but inter-related, dynamics. 

\section{Conclusions}\lb{con}

In this paper we have studied the Gompertz-Pareto distribution (GPD),
formed by the combination of the Gompertz curve, representing the
overwhelming majority of the economically less favorable part of the
population of a country, and the Pareto power law, describing its tiny
richest part. We discussed how the GPD is fully characterized by
only three positive parameters, inasmuch as boundary and continuity
conditions limit the parametric freedom of this distribution, and
which can be determined by linear data fitting. Equations for the
cumulative income distribution, complementary cumulative income
distribution, income probability density, Lorenz curve, Gini
coefficient and the percentage share of the Gompertzian part were
all written in this distribution. We discussed how the GPD allows
for an exponential approximation in its middle and upper sections
outside the Paretian region.

Application of this income distribution function was made to the 
Brazilian data from 1981 to 2005, previously published by Moura Jr.\
and Ribeiro \cite{nm09}, with additional new results for 2006 and 2007.
Consistency tests were carried out by comparing the Gini coefficients
obtained directly from the original data, without any assumption for
the shape and form of the distribution, with results obtained by using
the fitted parameters in order to re-obtain those coefficients. Similar
tests were made with the values of the percentage share of the Gompertzian
part of the distribution. The results indicate a general consistency
between the original values of both quantities as compared to the
calculated ones using the GPD parameters, although we found a
systematic, but mostly stable, discrepancy between these quantities
in the range of 6\% to 7\%. This small discrepancy might be due to some
statistical bias possibly present in the original calculation of the
GPD parameters of Brazil.

In conclusion, the results presented in this paper suggest that
the GPD does provide a coherent and analytically simple representation
for income distribution data leading to consistent results, at least
as far as data from Brazil is concerned.

\vspace{5mm}
We are grateful to 4 referees for their useful comments and suggestions,
as well as for pointing out various interesting papers which at the time
of writing the first version of this article we were unaware of.

\end{document}